\documentclass[pra,amsmath,amssymb,showpacs,superscriptaddress,twocolumn]{revtex4}

\usepackage{graphicx}
\usepackage{dcolumn}
\usepackage[hypertex]{hyperref}
\usepackage{bm}
\usepackage[usenames,dvipsnames]{color}
\usepackage{amsmath}
\usepackage{amsfonts}
\usepackage{amsthm}
\usepackage{amssymb}
\usepackage{mathrsfs}
\usepackage{subfigure}
\usepackage{ulem}

\begin{document}

\title{Multiple phase estimation for arbitrary pure states under white noise}

\author{Yao Yao}
\affiliation{Beijing Computational Science Research Center, Beijing, 100084, China}

\affiliation{Synergetic Innovation Center of Quantum Information and Quantum Physics, University of Science and Technology of China,
Hefei, Anhui 230026, China}

\author{Li Ge}
\affiliation{Beijing Computational Science Research Center, Beijing, 100084, China}

\author{Xing Xiao}
\affiliation{Beijing Computational Science Research Center, Beijing, 100084, China}

\author{Xiaoguang Wang}
\email{xgwang@zimp.zju.edu.cn}
\affiliation{Zhejiang Institute of Modern Physics, Department of Physics, Zhejiang University, Hangzhou 310027, China}

\author{C. P. Sun}
\email{cpsun@csrc.ac.cn}
\affiliation{Beijing Computational Science Research Center, Beijing, 100084, China}
\affiliation{Synergetic Innovation Center of Quantum Information and Quantum Physics, University of Science and Technology of China,
Hefei, Anhui 230026, China}

\date{\today}

\begin{abstract}
In any realistic quantum metrology scenarios, the ultimate precision in the estimation of parameters is limited not only by
the so-called Heisenberg scaling, but also the environmental noise encountered by the underlying system. In the context of
quantum estimation theory, it is of great significance to carefully evaluate the impact of a specific type of noise on
the corresponding quantum Fisher information (QFI) or quantum Fisher information matrix (QFIM). Here we investigate the
\textit{multiple} phase estimation problem for a natural parametrization of arbitrary pure states under \textit{white noise}. We
obtain the explicit expression of the symmetric logarithmic derivative (SLD) and hence the analytical formula of QFIM.
Moreover, the attainability of the quantum Cram\'{e}r-Rao bound (QCRB) is confirmed by the commutability of SLDs
and the optimal estimators are elucidated for the experimental purpose. These findings generalize previously known partial results
and highlight the role of white noise in quantum metrology.
\end{abstract}


\maketitle
%\section{INTRODUCTION}
Quantum metrology, emerged as a new branch of quantum technologies, provides a powerful and versatile framework for both theoretical and
experimental studies in the field of quantum-enhanced parameter estimation \cite{Giovanetti2004,Giovanetti2006,Pairs2009,Giovanetti2011}.
However, any realistic physical system will suffer from various environmental noises via the coupling with its surroundings \cite{Nielsen}.
As pointed out in Ref. \cite{Giovanetti2011}, analysis of the effects of noise is one of the major burgeoning trends of this field.
With the efforts of multiple authors, it is clearly evident that even a very low noise level can destroy the quadratic improvement over
the classical shot-noise limit \cite{Ji2008,Knysh2011,Escher2011,Demkowicz2011}. Although a unified method to deal with noise of arbitrary form
is still lacking, more in-depth study in this respect is continuing and the scope is far beyond the usual noisy quantum channels raised
in \cite{Escher2011,Demkowicz2011}. In fact, a plenty variety of significant physical effects or processes can also be regarded as the corresponding
noisy quantum channels in the context of quantum information theory. For instance, quite recently it is demonstrated that the relativistic effect
and quantum cloning machines are excellent platforms for investigating the quantum feature of quantum metrology scenarios
\cite{Hosler2013,Ahmadi2014,Yao2014a,Lu2013,Song2013,Yao2014b}.

On the other hand, due to the quantum Cram\'{e}r-Rao inequality, quantum Fisher information (QFI) is recognized as the key quantity to characterize the ultimate
precision in parameter estimation scenarios \cite{Helstrom1976,Holevo1982,Braunstein1994,Braunstein1996}. Therefore, a great amount of research
work of noisy quantum metrology can be translated into the evaluation of the dynamics of QFI in the presence of a certain kind of noise.
Though different kinds of upper bounds on QFI have been obtained for various purposes \cite{Escher2011,Demkowicz2011,Boixo2007,Escher2012,Alipour2014,Zwierz2014,Pang2014},
the analytical treatment of QFI is usually a difficult task. To summarize, we realize that all these analytical approaches in the literature can be classified
into the following three categories.

\textit{Method I.}  As shown by the seminal work of Braunstein and Caves, the QFI is intimately related to the the Bures distance or Uhlmann fidelity \cite{Braunstein1994}
\begin{align}
\mathcal{F}({\theta})&=4\lim_{\epsilon\rightarrow0}\left[\frac{\partial d_B(\rho_\theta,\rho_{\theta+\epsilon})}{\partial\epsilon}\right]^2 \nonumber\\
&=-2\lim_{\epsilon\rightarrow0}\frac{\partial^2 F_U(\rho_\theta,\rho_{\theta+\epsilon})}{\partial\epsilon^2},
\end{align}
where the Bures distance (and Uhlmann fidelity) between two quantum states $\rho_1,\rho_2$ can be defined as \cite{Bures1969,Uhlmann1976,Hubner1992}
\begin{align}
d_{B}(\rho_1,\rho_2)&=\sqrt{2-2\sqrt{F_U(\rho_1,\rho_2)}},\nonumber\\
F_U(\rho_1,\rho_2)&=\left(\textrm{Tr}\sqrt{\sqrt{\rho_1}\rho_2\sqrt{\rho_1}}\right)^2.
\end{align}
Therefore, instead of direct derivation, we can exploit the above relation to access the analytical formula of QFI if
one has already obtained the explicit expression of the Bures distance (or Uhlmann fidelity) of the corresponding states.
Actually, we recently notice that this strategy has already been applied successfully in several situations: the single qubit \cite{Zhong2013}, single-mode
Gaussian \cite{Pinel2013}, and two-mode Gaussian states \cite{Ahmadi2014,Adesso2014}.

\textit{Method II.} This strategy is based on the the spectral decomposition of the density operator $\rho(\theta)$
\begin{align}
\rho(\theta)=\sum_{k=1}^d\lambda_{k}(\theta)|\psi_k(\theta)\rangle\langle\psi_k(\theta)|,\label{decomposition}
\end{align}
where $d$ is the dimension of $\rho(\theta)$ and $\theta$ is the parameter to be estimated. Note that $\lambda_{k}(\theta)$ might be \textit{zero} for some $k$.
Using the Eq. (\ref{decomposition}) as the starting point, Pairs and O'Loan provided an explicit expression of QFI \cite{Pairs2009,O'Loan2014}
\begin{equation}
\mathcal{F}_\theta=\sum_{i=1}^d\frac{1}{\lambda_i}(\partial_\theta\lambda_i)^2
+2\sum_{j\neq k}^d\frac{(\lambda_j-\lambda_k)^2}{\lambda_j+\lambda_k}|\langle\psi_j|\partial_\theta\psi_k\rangle|^2,
\end{equation}
Consequently, Liu and Zhang \textit{et al.} went a step further, by noting that the symmetric logarithmic derivative (SLD) is only defined
on the support of $\rho(\theta)$. Therefore, the QFI can be rewritten as \cite{Liu2013,Zhang2013,Liu2014b}
\begin{align}
\mathcal{F}_\theta=&\sum_{i=1}^r\frac{1}{\lambda_i}(\partial_\theta\lambda_i)^2+\sum_{i=1}^r4\lambda_i\mathcal{F}_{Q,i} \nonumber\\
&-\sum_{j\neq k}^r\frac{8\lambda_j\lambda_k}{\lambda_j+\lambda_k}|\langle\psi_j|\partial_\theta\psi_k\rangle|^2,
\end{align}
where the QFI for pure eigenstate reads
\begin{equation}
\mathcal{F}_{Q,i}=4\left(\langle\partial_\theta\psi_i|\partial_\theta\psi_i\rangle-|\langle\psi_i|\partial_\theta\psi_i\rangle|^2\right).
\end{equation}
It should be emphasizing that now the summations go over the set $1\leq k \leq r$ and $r$ is the rank of $\rho(\theta)$.
The above expression is more convenient for the non-full-rank states and gives a clear physical meaning \cite{Liu2013,Zhang2013,Liu2014b}.
However, for arbitrarily high-dimensional states, it is not so easy to obtain a compact decomposition basis, especially when
the degeneracy of eigenvalues emerges \cite{Yao2014b}. Numerical studies may benefit more from this formula.

\textit{Method III.} The QFI is defined in terms of the SLD which satisfies the following the equation
\begin{equation}
\frac{\partial\rho_\theta}{\partial\theta}=\frac{1}{2}\left(\rho_\theta \mathcal{L}_{\theta}+ \mathcal{L}_{\theta} \rho_\theta\right).\label{SLD}
\end{equation}
If we obtain the explicit form of the Hermitian operator $\mathcal{L}_{\theta}$, then the calculation of QFI will be an easy task.
Nevertheless, the derivation of $\mathcal{L}_{\theta}$ is highly dependent on the structure of the density operator $\rho(\theta)$ and its
parametrization. For several special cases, the analytical solution of $\mathcal{L}_{\theta}$ can be found, involving some mathematical
tricks \cite{Collins2013,Monras2013,Knysh2013,Jiang2014}.

In this work, we thoroughly investigate the \textit{multiple} phase estimation problem for a natural parametrization of arbitrary pure states
under \textit{white noise}. The effect of white noise, also known as the (isotropic) depolarizing channel \cite{Nielsen} or \textit{Werner} state \cite{Werner1989},
is given by the map
\begin{equation}
|\Psi\rangle\langle\Psi|\Longrightarrow\rho^\textrm{w}=\eta|\Psi\rangle\langle\Psi|+\frac{1-\eta}{d}\mathbb{I}_{d\times d}.
\end{equation}
where $\eta$ is called the \textit{reliability} of the channel. This form of states has already played an essential role in various quantum information
tasks, such as quantum repeaters \cite{Briegel1998}, NMR quantum computing \cite{Braunstein1999} and quantum cloning machines \cite{Buzek1998}.
To accurately formulate the current problem and compare with the previous results \cite{Yao2014b}, we focus on the following parametrization
of arbitrary pure states
\begin{equation}
|\Psi(\bm{\phi})\rangle=\sum_{k=0}^{d-1}c_ke^{i\phi_k}|k\rangle.
\end{equation}
Without loss of generality, $c_k$ is assumed to be real since any imaginary part can be absorbed into the factor $e^{i\phi_k}$.
Now the parameter vector $\bm{\phi}=\{\phi_0,\phi_2,\ldots,\phi_{d-1}\}$ is the target to be inferred. However, the overall phase cannot be estimated,
so we can assume $\phi_0=0$. Here we adopt the \textit{Method III}, that is, trying to find out the analytical expression of the SLDs for $\rho^\textrm{w}$.

\textit{Calculation of SLD.} Before proceeding, the key observation which enables our calculation is that $\rho^\textrm{w}$ can be represented
in the exponential form
\begin{equation}
\rho^\textrm{w}=\eta\mathbb{P(\bm{\phi})}+\frac{1-\eta}{d}\mathbb{I}_{d\times d}=e^{\alpha\mathbb{P}(\bm{\phi})+\beta},
\end{equation}
by noting that
\begin{equation}
e^{\alpha\mathbb{P}(\bm{\phi})}=\mathbb{I}+(e^{\alpha}-1)\mathbb{P}(\bm{\phi}).
\end{equation}
For simplicity, here we define the von Neumann-type projector $\mathbb{P}(\bm{\phi})=|\Psi(\bm{\phi})\rangle\langle\Psi(\bm{\phi})|$.
Through direct calculation, we get the corresponding coefficients
\begin{equation}
\alpha=\ln\frac{(d-1)\eta+1}{1-\eta},\ \beta=\ln\frac{1-\eta}{d}.\label{coefficient}
\end{equation}

For states in the exponential form (e.g., $\rho(\theta)=e^{G(\theta)}$), Jiang provided a formal solution to the SLD \cite{Jiang2014}.
The derivation is based on two main observations. First, the derivative of $\rho(\theta)=e^{G(\theta)}$ can be cast
into an integral formula
\begin{equation}
\dot{\rho}=\int_0^1 e^{sG}\dot{G}e^{(1-s)G}ds.\label{derivative}
\end{equation}
where the overdot denotes the derivative with respect to $\theta$. Secondly, utilizing the Baker-Hausdorff formula, we have
\begin{equation}
e^GAe^{-G}=\sum_{n=0}^\infty\frac{1}{n!} G^{\times n}(A)=e^{G^{\times}}A,\label{Baker-Hausdorff}
\end{equation}
where the superoperator $G^{\times}$ is introduced and $G^{\times}$ denotes a commutator operation, namely \cite{Tanimura1989}
\begin{equation}
G^{\times}(A)=[G,A]=GA-AG.
\end{equation}
Combining Eqs. (\ref{SLD}), (\ref{derivative}) and (\ref{Baker-Hausdorff}), a formal expression of the SLD can be obtained \cite{Jiang2014}
\begin{equation}
\mathcal{L}=\sum_{n=0}^\infty f_n G^{\times n}(\dot{G})=f(G^{\times})(\dot{G}),\label{formula}
\end{equation}
where the generating function $f$ is determined by
\begin{equation}
f(t)=\sum_{n=0}^\infty f_n t^n=\frac{\tanh(t/2)}{t/2},
\end{equation}

To facilitate the solution of our problem, we define the following operator
\begin{equation}
\mathcal{A}_k=|\partial_{\phi_k}\Psi\rangle\langle\Psi|=ic_ke^{i\phi_k}|k\rangle\langle\Psi|,
\end{equation}
Thus the derivative of $\mathbb{P}$ with respect to $\phi_k$ is equal to
\begin{equation}
\dot{\mathbb{P}}_k=\mathcal{A}_k+\mathcal{A}_k^\dag.\label{projector}
\end{equation}
In addition, we have the following commutation relations
\begin{align}
\mathbb{P}^\times(\mathcal{A}_k)=[\mathbb{P}_k,\mathcal{A}_k]=ic_k^2\mathbb{P}-\mathcal{A}_k,\nonumber\\
\mathbb{P}^\times(\mathcal{A}_k^\dag)=[\mathbb{P}_k,\mathcal{A}_k^\dag]=ic_k^2\mathbb{P}+\mathcal{A}_k^\dag,
\end{align}
Intriguingly, we observe that the recursive structure of the nested-commutator appears
\begin{align}
\left[\mathbb{P}_k,[\mathbb{P}_k,\dot{\mathbb{P}}_k]\right]&=[\mathbb{P}_k,[\mathbb{P}_k,\mathcal{A}_k+\mathcal{A}_k^\dag]]\nonumber\\
&=-[\mathbb{P}_k,\mathcal{A}_k]+[\mathbb{P}_k,\mathcal{A}_k^\dag]\nonumber\\
&=\mathcal{A}_k+\mathcal{A}_k^\dag=\dot{\mathbb{P}}_k.\label{commutator}
\end{align}
With the help of the Taylor series expansion of the hyperbolic tangent function, the function $f$ can be rewritten as
\begin{equation}
f(t)=\sum_{n=0}^\infty\frac{4(4^{n+1}-1)B_{2n+2}}{(2n+2)!}t^{2n},
\end{equation}
where $B_{2n+2}$ is the $(2n+2)\textrm{th}$ Bernoulli number. It is remarkable that $f(t)$ only consists of the even-order terms,
which is compatible with the Hermiticity of the SLD.

Therefore, in our case, it is equivalent to define $G=\alpha\mathbb{P}(\bm{\phi})+\beta$. From the formula (\ref{formula}) and
the commutation relation (\ref{commutator}), we obtain the desired expression of the SLD
\begin{align}
\mathcal{L}_k=\alpha f(\alpha) \dot{\mathbb{P}}_k=2\tanh(\alpha/2)\dot{\mathbb{P}}_k,
\end{align}
From Eq. (\ref{coefficient}), we finally get
\begin{align}
\mathcal{L}_k=2\frac{e^\alpha-1}{e^\alpha+1}\dot{\mathbb{P}}_k=\frac{2d\eta}{2+(d-2)\eta}\dot{\mathbb{P}}_k.
\end{align}
Note that it is easy to check that $\textrm{Tr}(\rho^\textrm{w}\mathcal{L}_k)=0$,
since $\textrm{Tr}(\mathbb{P}\dot{\mathbb{P}}_k)=0$ and $\textrm{Tr}(\dot{\mathbb{P}}_k)=0$.

\textit{Evaluation of QFIM.} To compare with previous studies, here we analytically
evaluate the QFIM of $\rho^\textrm{w}(\bm{\phi})$ by use of $\mathcal{L}_k$. In the multi-parameter scenario, the
element of QFIM $\mathcal{F}(\bm{\phi})=[\mathcal{F}_{jk}]$ is defined by
\begin{equation}
\mathcal{F}_{jk}=\textrm{Tr}\left[\rho(\bm{\phi})\frac{\mathcal{L}_j \mathcal{L}_k+\mathcal{L}_k \mathcal{L}_j}{2}\right],
\end{equation}
where $\mathcal{L}_j$ and $\mathcal{L}_k$ are SLDs with respect to $\phi_j$ and $\phi_k$ respectively.
From Eq. (\ref{projector}), we have
\begin{align}
\dot{\mathbb{P}}_k^2=&c_k^2\left(|k\rangle\langle k|+|\Psi\rangle\langle\Psi|\right)\nonumber\\
&-c_k^3\left(e^{i\phi_k}|k\rangle\langle\Psi|+e^{-i\phi_k}|\Psi\rangle\langle k|\right),
\end{align}
Since $\rho^\textrm{w}(\bm{\phi})$ can be regarded as a mixture of $\mathbb{P}(\bm{\phi})$ and the identity operator $\mathbb{I}$,
the diagonal elements of QFIM is given by
\begin{align}
\mathcal{F}_{kk}=&\textrm{Tr}(\rho^\textrm{w}\mathcal{L}_k^2)\nonumber\\
=&(c_k^2-c_k^4)\left[x+\frac{2(1-\eta)}{d}\right]\left[\frac{2d\eta}{2+(d-2)\eta}\right]^2\nonumber\\
=&\frac{4d\eta^2}{2+(d-2)\eta}(c_k^2-c_k^4).
\end{align}
Correspondingly, the product $\dot{\mathbb{P}}_j\dot{\mathbb{P}}_k$ ($j\neq k$) takes a similar form
\begin{align}
\dot{\mathbb{P}}_j\dot{\mathbb{P}}_k=&-c_jc_k^2e^{i\phi_j}|j \rangle\langle \Psi|-c_j^2c_ke^{-i\phi_k}|\Psi \rangle\langle k|\nonumber\\
&+c_jc_ke^{i(\phi_j-\phi_k)}|j \rangle\langle k|,
\end{align}
Therefore, the off-diagonal elements of QFIM is given by
\begin{align}
\mathcal{F}_{jk}=&\bm{\mbox{Re}}\textrm{Tr}(\rho^\textrm{w}\mathcal{L}_j\mathcal{L}_k)\nonumber\\
=&-c_j^2c_k^2\left[x+\frac{2(1-\eta)}{d}\right]\left[\frac{2d\eta}{2+(d-2)\eta}\right]^2\nonumber\\
=&-\frac{4d\eta^2}{2+(d-2)\eta}c_j^2c_k^2 \quad (j\neq k).
\end{align}
Finally, the QFIM can be represented in a compact form
\begin{align}
\mathcal{F}_{jk}=\frac{4d\eta^2}{2+(d-2)\eta}\left(c_j^2\delta_{jk}-c_j^2c_k^2\right), \ \forall j,k.
\end{align}

Before moving forward, some remarks are in order. First, for the generalized $d$-dimensional equatorial pure states (e.g., $\eta=1$ and $c_k=1/\sqrt{d}$),
we recover the result of Ref. \cite{Humphreys2013}; meanwhile, if we only require that the amplitudes $c_k$ are equal, the result in \cite{Yao2014b}
is reestablished. Note that indeed the \textit{Method II} is employed in Ref. \cite{Yao2014b}, where a delicate choice of the decomposition basis
plays a critical role in the analysis. On the other hand, as a result of the monotonicity of QFI \cite{Fujiwara2001}, the following matrix inequality
should be satisfied
\begin{align}
\mathcal{F}\left(\rho^\textrm{w}(\bm{\phi})\right)\leq\eta\mathcal{F}\left(\mathbb{P}(\bm{\phi})\right),
\end{align}
where $\mathcal{F}(\mathbb{P}(\bm{\phi}))$ denotes the QFIM of the pure state $|\Psi(\bm{\phi})\rangle\langle\Psi(\bm{\phi})|$.
In fact, our result indicates that
\begin{align}
\mathcal{F}\left(\rho^\textrm{w}(\bm{\phi})\right)=\frac{d\eta^2}{2+(d-2)\eta}\mathcal{F}\left(\mathbb{P}(\bm{\phi})\right)\leq\eta\mathcal{F}\left(\mathbb{P}(\bm{\phi})\right).
\end{align}
Moreover, since the QFIM of $\rho^\textrm{w}(\bm{\phi})$ is proportional to that of $\mathbb{P}(\bm{\phi})$, we can define the \textit{ratio} function
\begin{equation}
\xi(\eta)\doteq\frac{d\eta^2}{2+(d-2)\eta}\leq 1.
\end{equation}
It is easy to check that $\xi(\eta)$ is a monotonically increasing function of the shrinking factor $\eta$.
This property of $\xi(\eta)$ confirms that (i) the QFI can \textit{never} be amplified in the presence of white noise;
(ii) the larger $\eta$ is, the more information $\rho^\textrm{w}(\bm{\phi})$ contains about parameters, which is
to be expected.

\textit{Attainability of QCRB.} In the multi-parameter scenarios, the celebrated quantum Cram\'{e}r-Rao bound (QCRB)
refers to the matrix inequality \cite{Helstrom1976}
\begin{equation}
\textrm{Cov}(\bm{\phi})\geq\big[M\mathcal{F(\bm{\phi})}\big]^{-1},\label{QCRB}
\end{equation}
where $\textrm{Cov}(\bm{\phi})$ stands for the covariance matrix of the unbiased estimator ${\bm{\hat\phi}}$
and $M$ is the number of measurements repeated ($M=1$ for definiteness). In sharp contrast to the single-parameter case, this lower bound
cannot be achieved in general, since simultaneous estimation of multiple parameters usually involves the joint
measurement of the corresponding \textit{incompatible} observables. The attainability problem of QCRB for pure states
has already been resolved by Fujiwara and Matsumoto \cite{Fujiwara2001a,Matsumoto2002}. For general mixed states, only recently
a series of research results by Gu\c{t}\u{a} \textit{et al.} reveal that the QCRB is asymptotically attainable
if and only if \cite{Guta}
\begin{equation}
\textrm{Tr}\left(\rho(\bm{\phi})[\mathcal{L}_j,\mathcal{L}_k]\right)=0\Leftrightarrow
\bm{\mbox{Im}}\textrm{Tr}(\rho(\bm{\phi})\mathcal{L}_j \mathcal{L}_k)=0,
\end{equation}
which is satisfied for all $j$ and $k$. In our study, the above calculation clearly shows that
\begin{equation}
\textrm{Tr}(\rho^\textrm{w}(\bm{\phi})\mathcal{L}_j \mathcal{L}_k)=\bm{\mbox{Re}}\textrm{Tr}(\rho^\textrm{w}(\bm{\phi})\mathcal{L}_j \mathcal{L}_k)\in\mathbb{R}.
\end{equation}
Therefore, the multi-parameter QCRB is achieveable in this particular case. Taking the trace of both sides of QCRB,
the total variance (error) of all the phases estimated follows the inequality
\begin{equation}
(\Delta\bm{\phi})^2=\sum_{\mu=1}^{d-1}(\Delta\phi_\mu)^2=\textrm{Tr}[\textrm{Cov}(\bm{\phi})]\geq \textrm{Tr}[\mathcal{F}(\bm{\phi})^{-1}].
\end{equation}
Note that this lower bound is also achievable due to the saturation of QCRB. In our case, it can be given as
\begin{align}
(\Delta\bm{\phi})^2_{\textrm{min}}=\sum_{\mu=1}^{d-1}F_\mu^{-1},
\end{align}
where $\{F_\mu\}_{\mu=1}^{d-1}$ is the set of eigenvalues of $\mathcal{F}(\rho^\textrm{w}(\bm{\phi}))$. Remember that
$\mathcal{F}(\rho^\textrm{w}(\bm{\phi}))$ is a $(d-1)\otimes (d-1)$ matrix. In addition, the possible symmetry of $|\Psi(\bm{\phi})\rangle$
may help us to access an analytical lower bound \cite{Yao2014b}.

To elucidate the optimal (joint) measurement of all the parameters, we follow the idea of Marzolino and Braun \cite{Marzolino2013} and \textit{generalize}
their method to our discussion. Based on the diagonalization of the inverse of QFIM, the QCRB (\ref{QCRB}) can be transformed into
\begin{equation}
\textrm{Cov}(\bm{\lambda})\geq\Lambda(\bm{\lambda})=\mathcal{Q}\mathcal{F(\bm{\phi})}^{-1}\mathcal{Q}^T,\label{QCRB1}
\end{equation}
where we define the column parameter vectors $\bm{\vec{\phi}}=\{\phi_k\}_{k=1}^{d-1}$, $\bm{\vec{\lambda}}=\{\lambda_k\}_{k=1}^{d-1}$
and $\bm{\vec{\lambda}}=\mathcal{Q}\bm{\vec{\phi}}$. Here $\mathcal{Q}$ is the \textit{orthogonal} matrix that diagonalizes $\mathcal{F(\bm{\phi})}^{-1}$
and the diagonal matrix $\Lambda(\bm{\lambda})=\bm{\mbox{diag}}\{F_{\lambda_1}^{-1},F_{\lambda_2}^{-1},\ldots,F_{\lambda_{d-1}}^{-1}\}$.
Since $\lambda_i=\mathcal{Q}_{ij}\phi_j$ (Einstein's summation convention) and $\mathcal{Q}$ is \textit{independent} of $\bm{\vec{\phi}}$
in our analysis, we have the equation
\begin{align}
\frac{\partial\rho}{\partial\phi_j}=&\frac{1}{2}(\mathcal{L}_{\phi_j}\rho+\rho\mathcal{L}_{\phi_j})=\frac{\partial\rho}{\partial\lambda_i}\mathcal{Q}_{ij}\nonumber\\
=&\frac{1}{2}(\mathcal{L}_{\lambda_i}\rho+\rho\mathcal{L}_{\lambda_i})\mathcal{Q}_{ij}.
\end{align}
Therefore, we have the relation $\mathcal{L}_{\phi_j}=\mathcal{Q}_{ij}\mathcal{L}_{\lambda_i}$, or equivalently,
$\mathcal{L}_{\bm{\vec{\phi}}}=\mathcal{Q}^T\mathcal{L}_{\bm{\vec{\lambda}}}$, where $\mathcal{L}_{\bm{\vec{\phi}}}$ and $\mathcal{L}_{\bm{\vec{\lambda}}}$
are the corresponding SLD vectors. Due to $\mathcal{Q}^T\mathcal{Q}=\mathbb{I}$, we finally arrive at $\mathcal{L}_{\bm{\vec{\lambda}}}=\mathcal{Q}\mathcal{L}_{\bm{\vec{\phi}}}$.
According to quantum estimation theory \cite{Pairs2009}, the optimal quantum estimator vector $\mathcal{O}_{\bm{\vec{\lambda}}}$ is given by
\begin{equation}
\mathcal{O}_{\lambda_k}=\lambda_k\mathbb{I}+\frac{\mathcal{L}_{\lambda_k}}{F_{\lambda_k}},\ (1\leq k\leq d-1)
\end{equation}
which attains the QCRB (\ref{QCRB1}) and achieves the desired property
\begin{equation}
\textrm{Cov}(\mathcal{O}_{\lambda_j},\mathcal{O}_{\lambda_k})=0, \  (j\neq k)
\end{equation}
implying that the optimal estimators for \textit{distinct} parameters are uncorrelated and the joint measurement can be realized.

In summary, we investigate the multiple phase estimation problem for a natural parametrization of arbitrary pure states under \textit{white noise}.
Our analysis extends and unifies several partial results for more specified states \cite{Yao2014b,Humphreys2013}. We have obtained
the analytical and compact expression of QFIM and also confirmed that the QCRB is universally attainable in this scenario.
Since the QFIM is irrespective of the parameters to be estimated, the lower bound of the total estimated error is proved to be a parameter-independent quantity,
which might be significant in other contexts. We also illustrate the optimal estimators attaining the QCRB for future experimental purpose.
Moreover, it is worth pointing out that our approach can be generalized to other circumstances, such as
\begin{equation}
\sigma=\frac{\eta}{r}\mathbb{\widetilde{P}}(\bm{\phi})+\frac{1-\eta}{d}\mathbb{I}_{d\times d},
\end{equation}
where the projection operator $\mathbb{\widetilde{P}}(\bm{\phi})$ is of the L\"{u}ders type \cite{Luders1951} and
$r$ is the rank of $\mathbb{\widetilde{P}}(\bm{\phi})$.

\textit{Acknowledgments.} This research is supported by the National Natural Science Foundation of China (Grants No. 11025527, No. 11121403, No. 10935010, No. 11074261, and No. 11247006),
the National 973 program (Grants No. 2012CB921602, No. 2012CB922104, and No. 2014CB921403), and the China Postdoctoral Science Foundation (Grant No. 2014M550598).

%%%%%%%%%%%%%%%%%%%%%%%%%%%%%%%%%%%%%%%%%%%%%%%%%%%%%%%%%%%%%%%%%%%%%%%%%%%%%%%%%%%%%%%%%%%%%

%%%%%%%%%%%%%%%%%%%%%%%%%%%%%%%%%%%%%%%%%%%%%%%%%%%%%%%%%%%%%%%%%%%%%%%%%%%%%%%%%%%%%%%%%%%%%

\end{document}